\documentclass[conference]{IEEEtran}
\usepackage{epsfig,rotating,setspace,latexsym,amsmath,epsf,amssymb,bm}
\usepackage{cite,graphicx,color,epstopdf,subcaption}
\usepackage{mathrsfs,multirow}
\usepackage{array,booktabs}

\begin{document}

\sloppy
\title{On the Degrees-of-freedom of the $3$-user MISO Broadcast Channel with Hybrid CSIT }

\author{
  \IEEEauthorblockN{SaiDhiraj Amuru}
  \IEEEauthorblockA{Department of ECE\\
    Virginia Tech, Blacksburg, VA\\
    \em{adhiraj@vt.edu} \vspace{-21pt}} 
  \and
  \IEEEauthorblockN{Ravi Tandon}
  \IEEEauthorblockA{Hume Center \& Department of ECE\\
    Virginia Tech, Blacksburg, VA\\
    \em{tandonr@vt.edu} \vspace{-21pt}} 
     \and
  \IEEEauthorblockN{Shlomo Shamai (Shitz)}
\IEEEauthorblockA{Department of Electrical Engineering\\ 
   The Technion,  Haifa, Israel\\
\em{sshlomo@ee.technion.ac.il } \vspace{-21pt}}
}

\newcommand{\SNR}{\mathsf{SNR}}
\newcommand{\PDD}{\mathsf{PDD}}
\newcommand{\PPD}{\mathsf{PPD}}
\newcommand{\DDD}{\mathsf{DDD}}
\newcommand{\PPP}{\mathsf{PPP}}
\newcommand{\DPP}{\mathsf{DPP}}
\newcommand{\DDP}{\mathsf{DDP}}
\newcommand{\DPD}{\mathsf{DPD}}
\newcommand{\PDP}{\mathsf{PDP}}

\newcommand{\PP}{\mathsf{PP}}
\newcommand{\DD}{\mathsf{DD}}
\newcommand{\PD}{\mathsf{PD}}

\newcommand{\Perfect}{\mathsf{P}}
\newcommand{\Delayed}{\mathsf{D}}
\newcommand{\None}{\mathsf{N}}

\newcommand{\CSI}{\mathsf{CSI}}
\newcommand{\CSIT}{\mathsf{CSIT}}
\newcommand{\JSIT}{\mathsf{JSIT}}

\newcommand{\dCSIT}{\mathsf{d-CSIT}}
\newcommand{\dJSIT}{\mathsf{d-JSIT}}

\newcommand{\CSIR}{\mathsf{CSIR}}
\newcommand{\JSIR}{\mathsf{JSIR}}

\newcommand{\JSI}{\mathsf{JSI}}
\newcommand{\DoF}{\mathsf{DoF}}
\newcommand{\AWGN}{\mathsf{AWGN}}

\newtheorem{Theo}{Theorem}
\newtheorem{remark}{Remark}
\newtheorem{Lem}{Lemma}
\newtheorem{Cor}{Corollary}
\newtheorem{Def}{Definition}
\vspace{-22pt}

\maketitle
\vspace{-20pt}

\begin{abstract}
The $3$-user multiple-input single-output (MISO) broadcast channel (BC) with hybrid channel state information at the transmitter $(\CSIT)$ is considered. In this framework, there is perfect and instantaneous $\CSIT$ from a subset of users and delayed $\CSIT$ from the remaining users.
We present new results on the degrees of freedom $(\DoF)$ of the $3$-user MISO BC with hybrid $\CSIT$. In particular, for the case of $2$ transmit antennas, we show that with perfect $\CSIT$ from one user and delayed $\CSIT$ from the remaining two users, the optimal $\DoF$ is $5/3$. For the case of $3$ transmit antennas and the same hybrid $\CSIT$ setting, it is shown that a higher $\DoF$ of $9/5$ is achievable and this result improves upon the best known bound. Furthermore, with $3$ transmit antennas, and the hybrid $\CSIT$ setting in which there is perfect $\CSIT$ from two users and delayed $\CSIT$ from the third one, a novel scheme is presented which achieves $9/4$ $\DoF$. Our results also reveal new insights on how to utilize hybrid channel knowledge for multi-user scenarios.

\end{abstract}
\vspace{0pt}
\section{Introduction}
\vspace{0pt}

There has been a significant recent interest in understanding the impact of delayed $\CSIT$ on the $\DoF$ of multi-user MIMO systems. Maddah-Ali and Tse \cite{MAT2012} showed that for the $K$-user MISO broadcast channel, with a $K$-antenna transmitter and $K$ single antenna users, the optimal sum $\DoF$ is given by the elegant formula $K/(1+ \frac{1}{2}+ \ldots + \frac{1}{K})$. This result shows that even completely delayed $\CSIT$ can significantly increase the $\DoF$ by exploiting overheard side-information at the users/receivers. However, this result assumes homogeneity in channel knowledge in the following sense: $\CSIT$ from every user is delayed.  This assumption may not always be true in practice and the delays experienced in acquiring $\CSIT$ can vary across users. Such scenarios can arise when some of the users can supply timely $\CSIT$ whereas others supply $\CSIT$ with delay (which could be a result of factors such as 
uplink overhead or infrequent feedback). This heterogeneity of channel knowledge motivates the framework of hybrid $\CSIT$. 

To formalize the hybrid $\CSIT$ framework, we denote the availability of $\CSIT$ from a particular receiver through a variable $I_{\CSIT}$, which can take values either $\Perfect$ or $\Delayed$. For receiver $k$, the state $I^k_{\CSIT}=\Perfect$ indicates that it supplies perfect and instantaneous $\CSIT$ and the state $I^k_{\CSIT}=\Delayed$ indicates that it supplies completely delayed $\CSIT$. 
Thus, for a $M$-antenna transmitter and $K$ single antenna receivers, i.e., the $(M, K)$ MISO BC, there are a total of $2^K$ possible $\CSIT$ configurations. The understanding of how to optimally utilize hybrid $\CSIT$ is far from complete and  optimal results are known only for the case of $(2,2)$ MISO BC. If the transmitter has perfect $\CSIT$ from both the receivers $(\PP)$, then the optimal $\DoF$ is $2$ which can be achieved using beamforming techniques \cite{MIMOBC}. When there is delayed $\CSIT$ from both the users $(\DD)$,   then the optimal $\DoF$ reduces to $4/3$ \cite{MAT2012}. For the hybrid $\CSIT$ scenario in which the transmitter has instantaneous $\CSI$ from receiver $1$ and delayed $\CSI$ from receiver $2$, (hybrid $\CSIT$: $\PD$) it was shown in \cite{Maleki_Retrospective} that the optimal $\DoF$ is $3/2$. 

We next come to the simplest non-trivial extension of the hybrid $\CSIT$ setting for more than $2$ receivers, i.e., the case of  three receiver, i.e. $(M, 3)$ MISO BC which is the main focus of this paper.  Here, a total of $2^{3} = 8$ possible $\CSIT$ configurations, namely $\PPP,\PPD,\PDP,\DPP,\PDD,\DPD,\DDP,\DDD$ can arise. Essentially, we have $4$ non-degenerate $\CSIT$ configurations, namely $\PPP,\PDD,\PPD,\DDD$, which depending on the number of transmit antennas ($M=2$ or $M=3$)  lead to two scenarios. 

The optimal $\DoF$ for the $(2,3)$ MISO BC is $2$ with either $\PPP$ or $\PPD$ configurations (limited by $2$  transmit antennas). 
For the $\DDD$ configuration, it has been shown in \cite{MAT2012} that the optimal $\DoF$ is given by $3/2$. Therefore, the only remaining case for which optimal $\DoF$ was not known prior to this work is the $\PDD$ configuration. In this paper, we show that the optimal $\DoF$ for this $\CSIT$ configuration is $5/3$. We next consider the $(3,3)$ MISO BC with hybrid $\CSIT$. For this setting, the optimal $\DoF$ is $3$ with the $\PPP$ configuration \cite{MIMOBC} and reduces to $18/11$ in the $\DDD$ configuration \cite{MAT2012}. However, the optimal $\DoF$ values for the remaining two $\CSIT$ configurations i.e., $\PDD$ and $\PPD$ are unknown. We present novel schemes for these two configurations which exploit hybrid channel knowledge to achieve sum $\DoF$ values of $9/5$ and $9/4$ respectively. The paper that is most relevant to this work is \cite{Mohanty2013}, in which outer bounds for the $(K, K)$ MISO BC are obtained for general hybrid $\CSIT$ configurations. In addition, a coding scheme for $(3,3)$ MISO BC with $\PDD$ configuration is given which achieves $5/3$ $\DoF$. Our results improve upon this bound to achieve $9/5$ $\DoF$, as well as establish that the optimal sum $\DoF$ for the $(2,3)$ MISO BC for $\PDD$ configuration is $5/3$. 

Our results show how to utilize hybrid $\CSIT$, which is a mixture of instantaneous and outdated $\CSIT$ from different receivers. While the core idea to exploit overheard side information at those receivers supplying delayed $\CSIT$ bears similarities to \cite{MAT2012}, the new technical challenge is that this exploitation must be done so that instantaneous $\CSIT$ (from other receivers) can be simultaneously harnessed.

\section{System Model}\label{system_model}
A $(M,K)$ MISO broadcast channel with $M$-transmit antennas and $K$-single 
antenna receivers with hybrid $\CSIT$ is considered. 
The received signal at the $k$th receiver is given by
\begin{align}\label{system_model_eq}
y_{k}(t)&= \mathbf{h}_{k}(t)\mathbf{x}(t) + z_{k}(t), 
\end{align}
where $\mathbf{x}(t)$  is the $M \times 1$ channel input at time $t$ with
$E\left(|\mathbf{x}(t)|^2\right)\leq P_T$, where $P_T$ is the average input power constraint; $\mathbf{h}_{k}(t)$ is the $1\times M$ channel vector from the transmitter to receiver $k$ at time $t$. Without loss of generality, 
$\mathbf{h}_k(t)$ is assumed to be sampled from any continuous 
distribution (e.g., Rayleigh) with an identity co-variance matrix, and are 
independent and identically distributed (i.i.d.) across time and also i.i.d.
across receivers. The additive noise 
$z_{k}(t)$ is distributed according to $\mathcal{CN}(0,1)$ for $k=1,\ldots,K$ and 
assumed to be independent of all other random variables. 
Throughout the paper, we assume the availability of global channel
state information at the receivers (i.e., full $\CSIR$). 

The rate tuple $(R_{1},R_{2},\ldots,R_K)$, with $R_{k}= \log(|W_{k}|)/n$, where 
$W_{k}$ is the message intended for the $k$th receiver, is achievable if 
there exist an encoding function and $K$ decoding functions (one for each receiver) such that the probability of 
decoding error at each receiver can be made arbitrarily small. 
The encoding function depends on the specific hybrid  $\CSIT$ configuration. 
For example, when the transmitter has perfect and instantaneous $\CSIT$ from the $1$st
receiver and delayed $\CSIT$ from the remaining $(K-1)$ receivers,
the encoding function depends on the current and past $\CSIT$ 
of the $1$st receiver and only the past $\CSIT$ of the other $(K-1)$ receivers.
For other hybrid $\CSIT$ settings, the encoding and the decoding functions
can be defined similarly. 
In this paper, we focus on the sum $\DoF$ of the $M$-antenna, $K$-receiver MISO BC, (henceforth referred to as the $(M,K)$-MISO BC) which is defined as $\textsf{DoF}(M,K) = \lim_{P_{T}\rightarrow \infty}
\max \sum_{k=1}^{K} \frac{R_{k}}{\log(P_{T})}$, where the maximum is over all achievable 
$K$-tuples $(R_{1}, \ldots, R_{K})$.  

\section{Main Results}\label{sec:Theorems}

\begin{Theo}\label{Theorem2111PDD}
The optimal sum $\DoF$ of the $(2,3)$ MISO BC with instantaneous $\CSIT$ from receiver $1$ and delayed $\CSIT$ 
from receivers $2$, $3$ i.e., $\left(I^1_{\CSIT},I^2_{\CSIT},I^3_{\CSIT}\right)=\PDD$ is
\begin{align}
\textsf{DoF}^{\PDD}(2,3)=\frac{5}{3}.
\end{align}
\end{Theo}

\begin{Theo}\label{Theorem3111PDD}
The sum $\DoF$ of the $(3,3)$ MISO BC with instantaneous $\CSIT$ from receiver $1$ and delayed $\CSIT$ 
from receivers $2$, $3$ i.e., $\left(I^1_{\CSIT},I^2_{\CSIT},I^3_{\CSIT}\right)=\PDD$ satisfies
\begin{align}
\frac{9}{5}\leq \textsf{DoF}^{\PDD}(3,3)\leq \frac{17}{9}.
\end{align}
\end{Theo}

\begin{Theo}\label{Theorem3111PPD}
The sum $\DoF$ of the $(3,3)$ MISO BC with instantaneous $\CSIT$ from receivers $1$ and $2$, and delayed $\CSIT$ 
from receiver $3$ i.e., $\left(I^1_{\CSIT},I^2_{\CSIT},I^3_{\CSIT}\right)=\PPD$ satisfies
\begin{align}
\frac{9}{4}\leq \textsf{DoF}^{\PPD}(3,3)\leq \frac{7}{3}.
\end{align}
\end{Theo}

\begin{figure}[t]
\centering
\includegraphics[width=1.0\columnwidth]{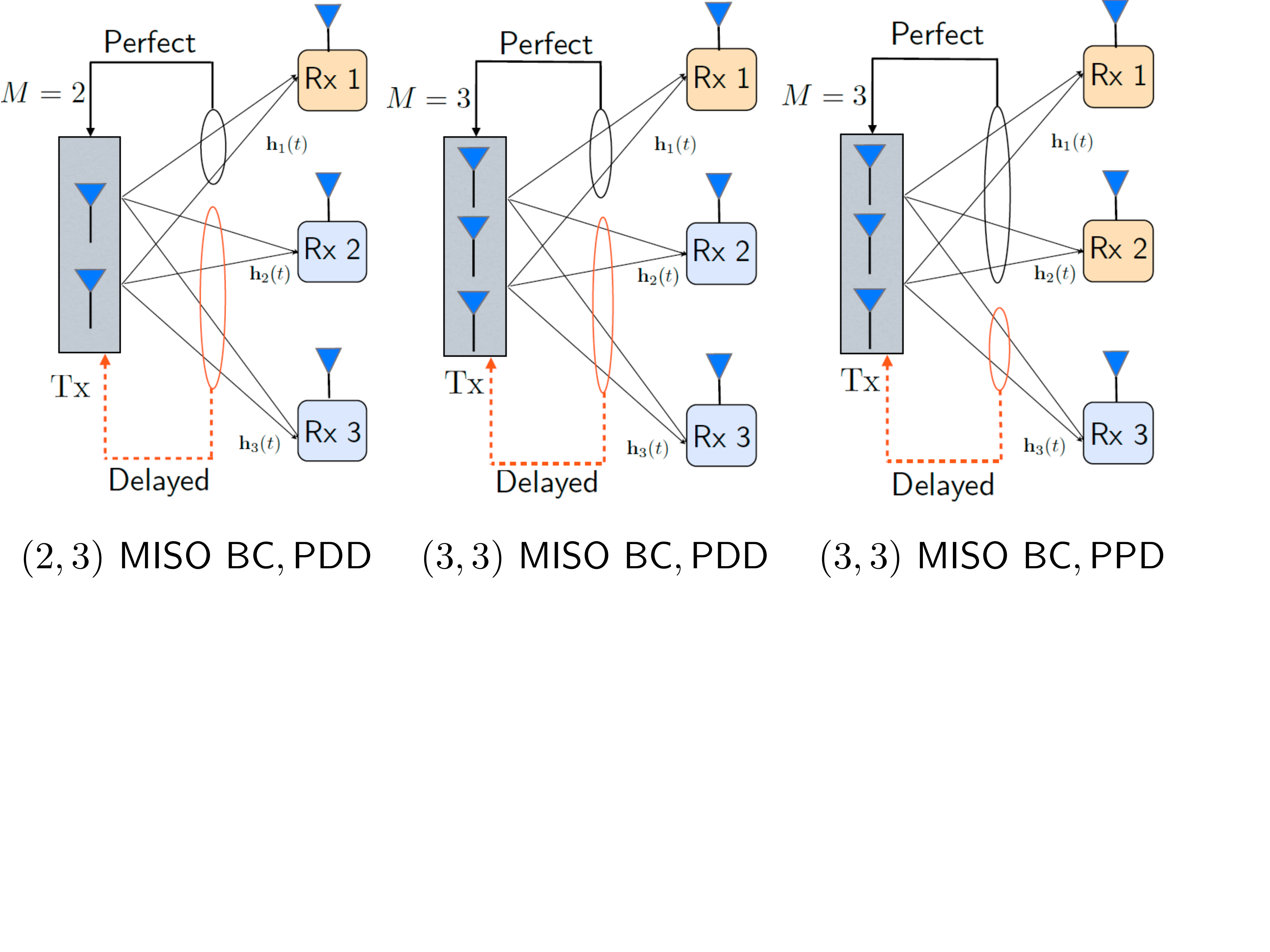}
\vspace{-79pt}
\caption{$(M, 3)$ MISO BC with Hybrid $\CSIT$.}
\label{fig:22MISOBC}
\vspace{-13pt}
\end{figure}

The converse proofs (upper bounds)  follow directly from the arguments in \cite{MAT2012,Mohanty2013,ElGamalFB} and are therefore omitted. 
Henceforth, we present new achievable schemes (lower bounds on $\DoF$) which are the main contributions of this paper. 

\vspace{-0pt}
\section{Achievability Proofs}\label{sec:Schemes}
\vspace{-0pt}

 We first refresh the optimal scheme for the $(2,2)$ MISO BC  \cite{Maleki_Retrospective} with hybrid $\CSIT$ configuration $\PD$ (instantaneous $\CSIT$ from user $1$, delayed $\CSIT$ from user $2$) that achieves the $\DoF$ value of $3/2$. Here, the transmitter sends two symbols $(a_{1}, a_{2})$ to user $1$ and one symbol $b$ to user $2$ in two time slots as follows: in the first time slot, it sends $[a_{1} \ \ a_{2}]^{T}+ \mathbf{h}_{1}(1)^{\perp}[b \ \ 0]^{T}$, where $\mathbf{h}_{1}(1)^{\perp}$ denotes a $2\times 2$ projection operator orthogonal to $\mathbf{h}_{1}(1)$. User $1$ receives a linear combination $\mathcal{L}_{1}(a_{1}, a_{2})$, whereas user $2$ gets a linear combination of $(a_{1}, a_{2})$ and $b$, denoted by $\mathcal{L}_{2}(a_{1}, a_{2}) + b$. Thus, $\mathcal{L}_{2}(a_{1}, a_{2})$ is a symbol which is desirable by both users since user $1$ can decode $(a_{1}, a_{2})$ from $\mathcal{L}_{1}(a_{1}, a_{2}), \mathcal{L}_{2}(a_{1}, a_{2})$, whereas user $2$ can use it to decode $b$. Thus, we call $\mathcal{L}_{2}(a_{1}, a_{2})$ as an {\it{order $2$ symbol}}, i.e., a symbol desired by $2$ users; this symbol can be sent in the second time slot achieving $3/2$ $\DoF$.

\vspace{-4pt}
\subsection{Order $2$ $\DoF$: $(2,3)$ MISO BC $(\PDD)$}
While the order $2$ $\DoF$ in the $(2,2)$ MISO BC is $1$ (one order $2$ symbol can be delivered to two receivers in one time slot), one can do better when we consider the extension  to the $3$-user MISO BC.  For the $3$-user MISO BC, there are $3$ possible types of order $2$ symbols, namely, symbols desired by receivers $(1,2)$,
symbols desired by receivers $(1,3)$, and symbols desired by receivers $(2,3)$. 
 We first present the optimal degrees of freedom region for delivering order $2$ symbols with hybrid $\CSIT$.  This result forms the basis for establishing the achievability proofs for Theorems~\ref{Theorem2111PDD} and \ref{Theorem3111PDD}. 

\begin{figure}[t]
\centering
\includegraphics[width=1.0\columnwidth]{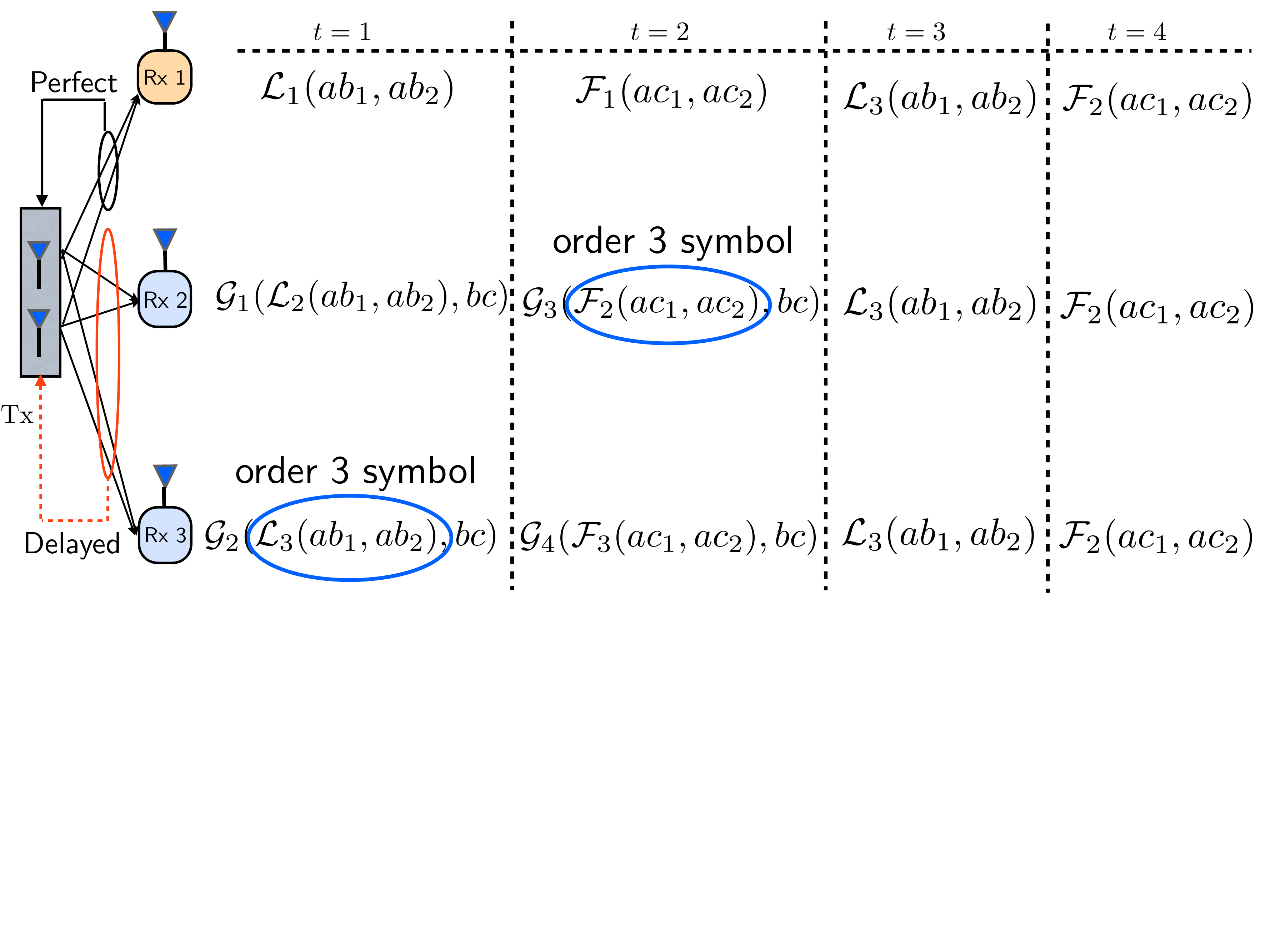}
\vspace{-76pt}
\caption{Achieving $5/4$ order $2$ $\DoF$; $(2,3)$ MISO BC.}
\label{fig:2111_Stage2}
\vspace{-15pt}
\end{figure}

\begin{Lem}\label{Lemma1}
The order $2$ $\DoF$ region for the $(2,3)$ MISO BC with $\PDD$ $\CSIT$ configuration
is given by
\begin{align}
d_{12}+d_{13}&\leq 1\label{eq1DoF2} \\
2(d_{12}+d_{23})+d_{13}&\leq 2\\
d_{12}+2(d_{23}+d_{13})&\leq 2. \label{eq5DoF2}
\end{align}
\end{Lem}
\textit{Proof:} The converse proof follows from the arguments similar in \cite{MAT2012,Mohanty2013,ElGamalFB} and is therefore omitted. From \eqref{eq1DoF2}-\eqref{eq5DoF2}, the optimal order $2$ sum $\DoF$ is $5/4$ corresponding to the tuple $(d_{12},d_{23},d_{13})=\left(\frac{1}{2},\frac{1}{4},\frac{1}{2}\right)$. We next present a novel coding scheme which achieves this tuple. To this end, we denote $ab,bc$ and $ac$-symbols as the order $2$ symbols desired by receivers $(1,2)$, $(2,3)$ and $(1,3)$ respectively and present a scheme which sends $(ab_1, ab_2, ac_1, ac_2, bc)$ to the corresponding receivers in $4$ time slots. 
This scheme is shown in Fig.~\ref{fig:2111_Stage2} and is described next: 

\noindent $\bullet$ At $t=1$,  transmitter sends $\{a b_1,ab_2\}$ along with $bc$ in a direction orthogonal
to $\mathbf{h}_1(1)$ as $\mathbf{x}(1)=[ab_1 \ ab_2]^T+\mathbf{h}_1^{\perp}(1)[bc \ 0 ]^{T}$. The receiver outputs are shown in Fig.~\ref{fig:2111_Stage2}, where $\mathcal{G}_i(\mathcal{L}_j(ab_1,ab_2),bc)$ indicates a linear combination of 
$\mathcal{L}_j(ab_1,ab_2)$ and $bc$.  At the end of $t=1$, we note that the linear combination $\mathcal{L}_3(ab_1,ab_2)$ is useful for all the $3$ receivers, i.e., this is an order 3 symbol. 

\noindent $\bullet$ At $t=2$, an identical scenario is created with the $ac$ symbols by sending
$\mathbf{x}(2)=[ac_1 \ ac_2]^T+\mathbf{h}_1^{\perp}(2)[bc \ 0]^{T}$. The corresponding  
outputs at the receivers are shown in Fig.~\ref{fig:2111_Stage2}, 
and similar to $\mathcal{L}_3(ab_1,ab_2)$, $\mathcal{F}_2(ac_1,ac_2)$ is also an order 3 symbol . 

\noindent $\bullet$ We now note that these order $3$ symbols, i.e., $\mathcal{L}_3(ab_1,ab_2)$ and $\mathcal{F}_2(ac_1,ac_2)$ can be reconstructed at the transmitter via delayed $\CSIT$ from users $2$ and $3$. These symbols can be subsequently delivered in time slots $t=3$ and $t=4$ (as order $3$ $\DoF$ for a $3$-receiver MISO BC is $1$). 

Finally, at the end of these $4$ time slots, the $1$st receiver can decode $ab_1,ab_2$ using 
$\mathcal{L}_1(ab_1,ab_2),\mathcal{L}_3(ab_1,ab_2)$ and
$ac_1,ac_2$ using $\mathcal{F}_1(ac_1,ac_2),\mathcal{F}_2(ac_1,ac_2)$.
The $2$nd receiver can decode $bc$ by canceling the interference from $\mathcal{F}_2(ac_1,ac_2)$. This $bc$ symbol is used to recover $\mathcal{L}_2(a b_1,a b_2)$ from $\mathcal{G}_1(\mathcal{L}_2(ab_1,ab_2),bc)$. Thus using $\mathcal{L}_2(a b_1,a b_2), \mathcal{L}_3(a b_1,a b_2)$, it can reconstruct the symbols $ab_1,ab_2$ by solving two LCs of two symbols. Along similar lines, the $3$rd receiver can reconstruct $(ac_{1}, ac_{2}, bc)$. 
Thus, the order $2$ $\DoF$ tuple $(d_{12},d_{23},d_{13})=\left(\frac{1}{2},\frac{1}{4},\frac{1}{2}\right)$ is achievable for $(2,3)$ MISO BC.

\vspace{3pt}
\subsection{Theorem~\ref{Theorem2111PDD}: $5/3$ $\DoF$ $-$ $(2,3)$ MISO BC $(\PDD)$}
We next present the optimal scheme that uses the result of Lemma~\ref{Lemma1} to achieve the $\DoF$ tuple $(d_{1}, d_{2}, d_{3})= \left(1, 1/3,1/3\right)$. Specifically, the scheme sends $12$ symbols to the $1$st receiver and $4$ symbols each to receivers $2$ and $3$ (denoted by $\{a_i\}_{i=1}^{12}$, $\{b_i\}_{i=1}^4$ and $\{c_i\}_{i=1}^4$)  in $12$ time slots. 

{\underline{Stage 1 -- generating order $2$ symbols:}}
This stage consists of $3$ phases, denoted by phase-$bc$, phase-$ab$ and phase-$ac$. Phase-$bc$ corresponds to creating one order $2$ symbol for receivers $(2,3)$. Phase-$ab$ and phase-$ac$ are used to create $\{ab_1,ab_2\}$ and $\{ac_1,ac_2\}$ respectively. Fig.~\ref{fig:2111_Stage1} shows the outputs at the receivers in this stage and the mechanism of generating order $2$ 
symbols. 

\noindent {\textit {Phase-$bc$: creating $1$ $bc$ symbol}}

 $\bullet$ At $t=1$, the transmitted and received signals are
given by
\begin{align}
\hspace{-5pt}\mathbf{x}(1)\hspace{-2pt}=\hspace{-2pt}\begin{bmatrix}a_1\\a_2\end{bmatrix}\hspace{-2pt}+\hspace{-2pt}\mathbf{h}_1^{\perp}(1)\begin{bmatrix}b_1\\0 \end{bmatrix}
\rightarrow\hspace{-2pt}
\begin{bmatrix}
y_1(1)\\
y_2(1)\\
y_3(1)
\end{bmatrix}\hspace{-2pt}=\hspace{-2pt}
\begin{bmatrix}
\mathcal{A}_1 \\
\mathcal{L}_1(\mathcal{A}_2,b_1)  \\
\mathcal{L}_2(\mathcal{A}_{1,2},b_1)
\end{bmatrix}.
\end{align} 
At the end of $t=1$, notice that $\mathcal{A}_2$ is useful for both $1$st and $2$nd receivers. 
Since the transmitter has delayed $\mathsf{CSI}$ from the $2$nd receiver, it can reconstruct $\mathcal{A}_2$. Note that $\mathcal{A}_1, \mathcal{A}_2$ and $\mathcal{A}_{1,2}$ are linear combinations of the two symbols $(a_1,a_2)$.

 $\bullet$ At $t=2$, the transmitter sends $\mathcal{A}_2$ along with a new symbol $b_2$ as $\mathbf{x}(2)=[\mathcal{A}_2 \ 0]^T+\mathbf{h}_1^{\perp}(2)[b_2 \ 0]^{T}$ and the outputs at the receivers are shown in Fig.~\ref{fig:2111_Stage1}.
At the end of $t=2$, the $1$st receiver can decode the symbols
$a_1,a_2$ using two linearly independent combinations $\mathcal{A}_1$ and $\mathcal{A}_2$.
$\mathcal{L}_4(\mathcal{A}_2,b_2)$ (with receiver $3$) is useful for the $2$nd receiver as it helps decode $b_1,b_2$ (since it can recover $3$ symbols $\mathcal{A}_2,b_1,b_2$ from $\mathcal{L}_1,\mathcal{L}_3$ and $\mathcal{L}_4$). Thus, $\mathcal{L}_4(\mathcal{A}_2,b_2)$ is an ingredient for creating the $bc$-symbol.

 $\bullet$ At $t=3$ and $t=4$, the transmitter sends new symbols $a_3,a_4$ for the $1$st receiver along with $c_1,c_2$ for the $3$rd receiver as $\mathbf{x}(3)=[a_3\ a_4]^T+\mathbf{h}_1^{\perp}(3)[c_1 \ 0]^{T}$ and 
$\mathbf{x}(4)=[\mathcal{A}_4 \ 0]^T+\mathbf{h}_1^{\perp}(2)[c_2 \ 0]^{T}$,
and the outputs at the receivers are shown in Fig.~\ref{fig:2111_Stage1}. The side information $\mathcal{L}_7(\mathcal{A}_4,c_2)$ at receiver $2$ is useful for the $3$rd receiver (to recover $c_1,c_2$). Therefore from $t=2$ and $t=4$, $\mathcal{L}_4(\mathcal{A}_2,b_2) + \mathcal{L}_7(\mathcal{A}_4,c_2)$ is the $bc$ symbol. 

\begin{figure}[t]
\centering
\vspace{5pt}
\includegraphics[width=1.0\columnwidth, height=210pt, keepaspectratio=true]{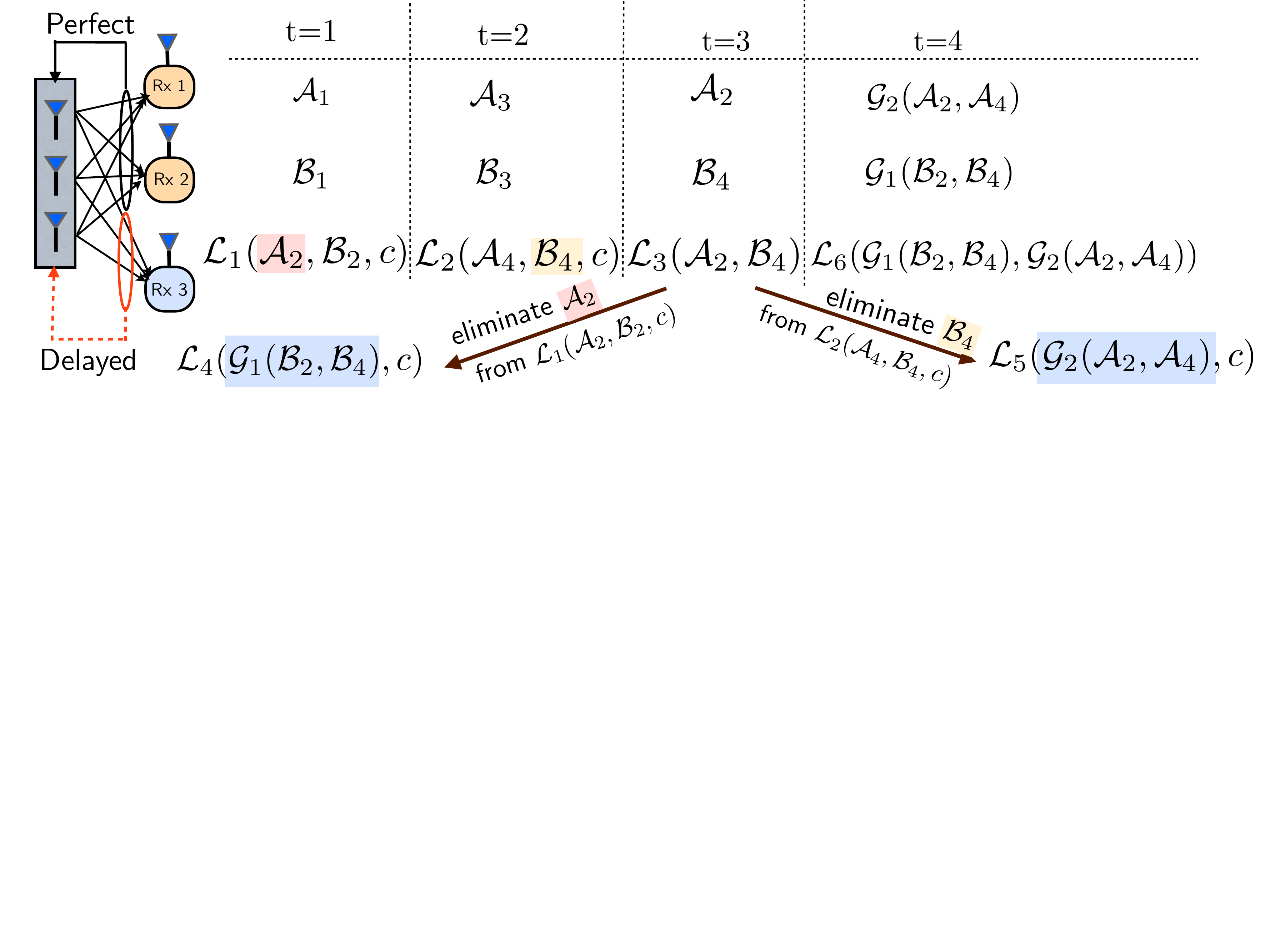}
\vspace{-113pt}
\caption{Achieving $9/4$ $\DoF$ for $(3,3)$ MISO BC  with $\PPD$.}
\label{fig:3111Stage1_Phase1_PPD}
\vspace{-15pt}
\end{figure}

\noindent {\textit {Phase-$ab$: creating $2$ $ab$ symbols}

Here, we create $2$ $ab$-symbols. In this regard, the transmitter sends $\{a_j\}_{j=5}^8$ to the $1$st receiver and $(b_3,b_4)$ to the $2$nd receiver as $\mathbf{x}(5)=[a_5 \ a_6]^T+\mathbf{h}_1^{\perp}(5)[b_3 \ 0]^{T}$,
$\mathbf{x}(6)=[a_7 \ a_8]^T+\mathbf{h}_1^{\perp}(6)[b_4 \ 0]^{T}$. It is clear from  Fig.~\ref{fig:2111_Stage1}, that $\mathcal{A}_{6}$ and $\mathcal{A}_{8}$ are two $ab$-symbols useful for both the receivers $1$ and $2$. 

\noindent {\textit {Phase-$ac$: creating $2$ $ac$ symbols}

The above phase is repeated here by sending $\mathbf{x}(7)=[a_9 \ a_{10}]^T+\mathbf{h}_1^{\perp}(7)[c_3 \ 0]^{T}$ and $\mathbf{x}(8)=[a_{11} \  a_{12}]^T+\mathbf{h}_1^{\perp}(8)[c_4 \ 0]^{T}$. It is clear from Fig.~\ref{fig:2111_Stage1} that $\mathcal{A}_{10}$ and $\mathcal{A}_{12}$ are $ac$-symbols desired by receivers $1$ and $3$. 

 Summary of Stage $1$: At the end of stage $1$, we have $5$ order $2$ symbols: $(ab_1, ab_{2}) = (A_6, A_8)$, $(ac_1, ac_2) = (A_{10}, A_{12})$ and $bc=\mathcal{L}_4(\mathcal{A}_2,b_2)+\mathcal{L}_7(\mathcal{A}_4,c_2)$ that can be delivered in stage $2$. 

{\underline{Stage 2 -- delivering order $2$ symbols:}}
The $5$ order 2 symbols created in stage $1$ can be delivered in $4$ time slots
using the scheme developed in Lemma~\ref{Lemma1}. Upon receiving these order $2$ symbols, 
all receivers can decode their desired symbols. For example, upon receiving 
$\mathcal{A}_6$, the $1$st receiver can decode $a_5,a_6$ via using $\mathcal{A}_5$ which was already received at $t=5$. The $2$nd receiver can use $\mathcal{A}_6$ to cancel interference in the symbol $\mathcal{L}_9(\mathcal{A}_6,b_3)$ and decode $b_3$. Similar reasoning holds true for the other symbols at the receivers. 
Overall, the transmitter spent $8$ time slots in the $1$st stage and $4$ time slots in the $2$nd stage, which gives the optimal $\DoF$ tuple $\left(\frac{12}{12},\frac{4}{12},\frac{4}{12}\right)=\left(1,\frac{1}{3},\frac{1}{3}\right)$ i.e., $\DoF^{\PDD}(2,3)=5/3$. 
\begin{figure*}[t]
\centering
\includegraphics[width=0.76\textwidth, keepaspectratio=true]{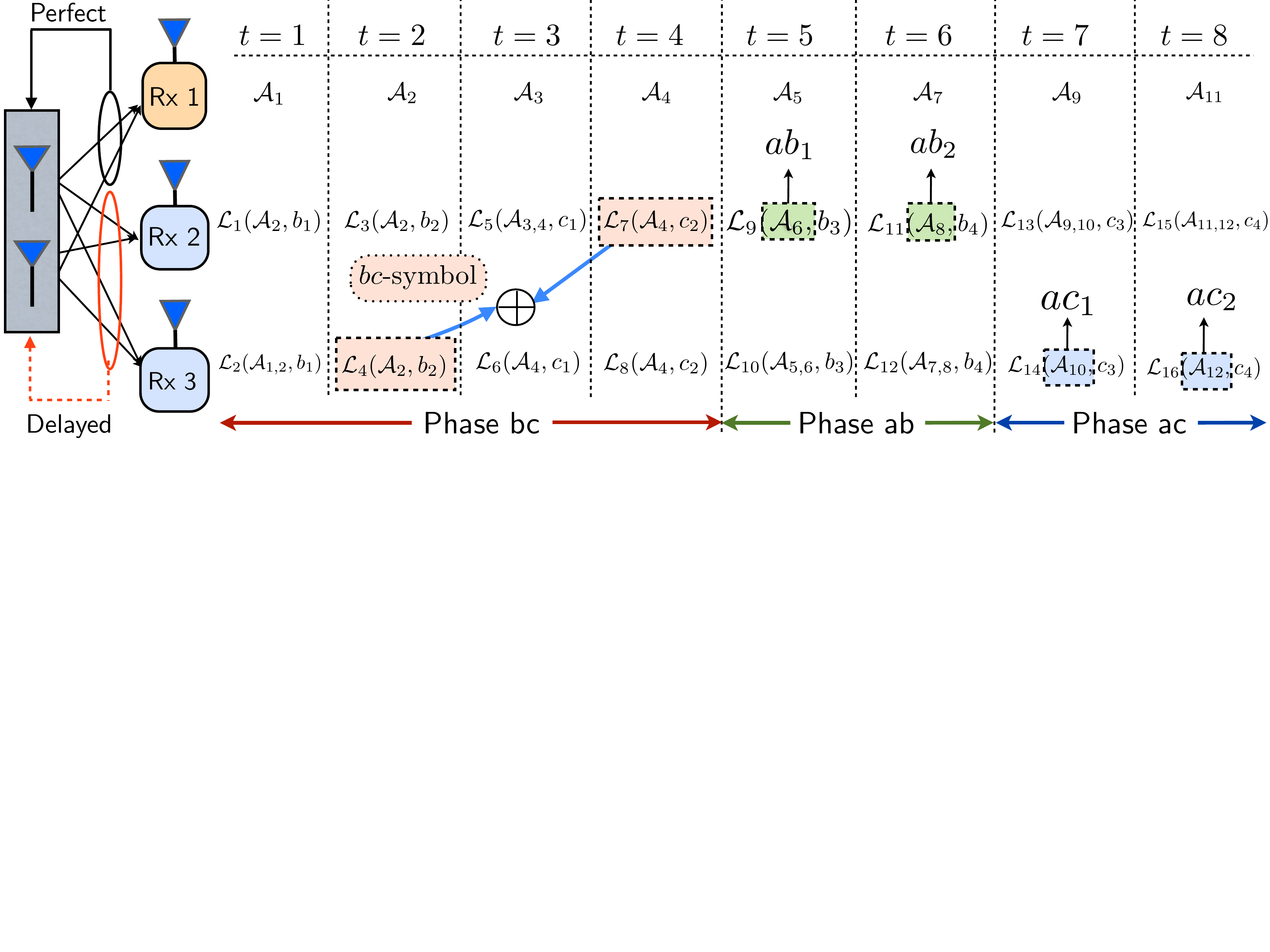}
\vspace{-154pt}
\caption{Generating order $2$ symbols -- $(2,3)$ MISO BC, $\CSIT$ configuration $\PDD$.}
\label{fig:2111_Stage1}
\vspace{-15pt}
\end{figure*}

\subsection{Theorem~\ref{Theorem3111PDD}:  $9/5$ $\DoF$ $-$ $(3,3)$ MISO BC $(\PDD)$}} \label{Sec:17by9}

In this section, we present a scheme that achieves the tuple $(d_{1}, d_{2}, d_{3})= \left(1, \frac{2}{5}, \frac{2}{5}\right)$, i.e., total of $9/5$ $\DoF$ and improves upon the best known bound of $5/3$ \cite{Mohanty2013}. We present a scheme that sends $10$ symbols to the $1$st receiver and $4$ symbols each to receivers $2$ and $3$ in a total of $10$ time slots.

\begin{remark}
Similar to the scheme for $(2,3)$ MISO BC, this scheme also has two stages, stage $1$ dedicated to generating order $2$ symbols and stage $2$ for their delivery  using Lemma~\ref{Lemma1}. However, there are two key distinctions --  a) the mechanism of generating order $2$ symbols is different, and b) the rate of creation of order $2$ symbols is higher for $(3,3)$ MISO BC, leading to a higher $\DoF$ value of $9/5$ compared to $5/3$. 
\end{remark}

{\underline{Stage 1 -- generating order $2$ symbols:}}
This stage (shown in Fig.~\ref{fig:3111_AllStages}) is split into two distinct phases: 
Phase-$bc$ takes $3$ time slots and generates $1$ $bc$-symbol and 
Phase-$(ab, ac)$ takes $3$ time slots to {\textit{jointly}} generate $2$ $ab$-symbols and $2$ $ac$-symbols. 

\noindent {\textit {Phase-$bc$: creating $1$ $bc$-symbol}}

This phase sends $(a_{1}, a_{2}, a_{3})$ along with $(b_{1}, b_{2})$ and $(c_{1}, c_{2})$ in three time slots as follows:
\begin{alignat*}{3}
\mathbf{x}(1)&=
\begin{bmatrix}a_1\\a_2\\a_3
\end{bmatrix}\ & & \rightarrow 
\begin{bmatrix}
y_1(1)\\
y_2(1)\\
y_3(1)
\end{bmatrix}
=\begin{bmatrix}
\mathcal{A}_1\\
\mathcal{A}_2\\
\mathcal{A}_3
\end{bmatrix}, 
\nonumber \\
\mathbf{x}(2)&=
\begin{bmatrix}
0\\
\mathcal{A}_2\\
0
\end{bmatrix}+ \ &\mathbf{h}_1^{\perp}(2)
\begin{bmatrix}0\\ b_{1}\\b_2
\end{bmatrix} & \rightarrow
\begin{bmatrix}
y_1(2)\\
y_2(2)\\
y_3(2)
\end{bmatrix}
=\begin{bmatrix}
\mathcal{A}_2\\
\mathcal{L}_1(\mathcal{A}_2,\mathcal{B}_1)\\
\mathcal{L}_2(\mathcal{A}_2,\mathcal{B}_2)\\
\end{bmatrix}, \nonumber\\
\mathbf{x}(3)&=
\begin{bmatrix}
0\\
0\\
\mathcal{A}_3
\end{bmatrix} + \ & \mathbf{h}_1^{\perp}(3)
\begin{bmatrix}0\\ c_{1}\\c_2
\end{bmatrix} & \rightarrow
\begin{bmatrix}
y_1(3)\\
y_2(3)\\
y_3(3)
\end{bmatrix}
=\begin{bmatrix}
\mathcal{A}_3\\
\mathcal{G}_1(\mathcal{A}_3,\mathcal{C}_1)\\
\mathcal{G}_2(\mathcal{A}_3,\mathcal{C}_2)\\
\end{bmatrix}. \nonumber
\end{alignat*}
It is clear that at the end of this phase, receiver $1$ is able to decode $(a_{1},a_2,a_3)$ and the transmitter can create the following $bc$-symbol that is useful for both the receivers $2$ and $3$: \break $bc= \mathcal{L}_2(\mathcal{A}_2,\mathcal{B}_2)+\mathcal{G}_1(\mathcal{A}_3,\mathcal{C}_1)$.

\noindent {\textit {Phase-$(ab,ac)$: creating $2$ $ab$-symbols and $2$ $ac$-symbols}}

 This phase sends $\{a_{j}\}_{j=4}^{10}$ for receiver $1$, $(b_{3}, b_{4})$ for receiver $2$ and
$(c_{3}, c_{4})$ for receiver $3$ to generate $2$ $ab$ and $2$ $ac$-symbols in $3$ time slots (at a higher rate in comparison to Theorem \ref{Theorem2111PDD}).

 $\bullet$ At $t=4,5$, the transmitter sends $\mathbf{x}(4)=[a_4 \ a_5 \ a_6]^T+\mathbf{h}_1^{\perp}(4)[0 \ b_3 \ b_4]^T$ and  $\mathbf{x}(5)=[a_7 \ a_8 \ a_9]^T+\mathbf{h}_1^{\perp}(5)[0 \ c_3 \ c_4]^T$. From Fig.~\ref{fig:3111_AllStages}, note that  
$ \mathcal{L}_4(\mathcal{A}_6,\mathcal{B}_4)$ (at user $3$) and $\mathcal{G}_3(\mathcal{A}_{8},\mathcal{C}_3)$ (at user $2$) are useful for users $2$ and $3$ respectively. 

 $\bullet$ Thus at $t=6$, the transmitter sends these side information symbols along with $a_{10}$, a new symbol for receiver $1$ as $\mathbf{x}(6)=[a_{10} \ 0 \ 0]^T+\mathbf{h}_1^{\perp}(6)[0 \ \mathcal{L}_4(\mathcal{A}_6,\mathcal{B}_4) \ \mathcal{G}_3(\mathcal{A}_{8},\mathcal{C}_3)]^T$.

We next note that in Phase-$(ab,ac)$, receiver $1$ has obtained a total of $3$ interference free symbols and requires $4$ more useful symbols in order to decode $\{a_{j}\}_{j=4}^{10}$.  Receiver $2$ uses $y_{2}(5)$ and
$y_{2}(6)$ to eliminate  $\mathcal{G}_3(\mathcal{A}_{8},\mathcal{C}_3)$ to obtain a LC of $(a_{10}, \mathcal{A}_{6})$ and $\mathcal{B}_{4}$ denoted as $\mathcal{A}^{'}_{6} + \mathcal{B}_{4}$. Receiver $2$ also obtains $\mathcal{L}_3(\mathcal{A}_5,\mathcal{B}_3)$ at $t=4$ which leads to the generation of  $2$ $ab$-symbols: $\mathcal{A}_5$ and $\mathcal{A}^{'}_{6}$.  Receiver $3$ uses $y_{3}(4)$ and $y_{3}(6)$ to eliminate  $\mathcal{L}_4(\mathcal{A}_6,\mathcal{B}_4)$ to obtain a LC of $(a_{10}, \mathcal{A}_{8})$ and $\mathcal{C}_{3}$ denoted as $\mathcal{A}^{'}_{8} + \mathcal{C}_{3}$. 
Thus $\mathcal{A}_{9}$ and $\mathcal{A}^{'}_{8}$ are $2$ two $ac$ symbols. 

{\underline{Stage 2 -- delivering order $2$ symbols:}}
This stage delivers the $5$ order $2$ symbols created in stage $1$, i.e., $(ab_1, ab_2, ac_1, ac_2, bc)$ in $4$ time slots using the transmission scheme of Lemma~\ref{Lemma1} (also a valid scheme for the $(3,3)$ MISO BC). 
Overall we delivered $10$ symbols to the $1$st receiver, $4$ symbols each to the $2$nd and $3$rd receivers in $10$ time slots. Thus the $\DoF$ achieved by this scheme is given by
\begin{align}
\DoF^{\PDD}(3,3)\geq \frac{10+4+4}{3+3 +4}=\frac{9}{5},
\end{align}
i.e., the $\DoF$ triplet $\left(\frac{10}{10},\frac{4}{10},\frac{4}{10}\right)=\left(1,\frac{2}{5},\frac{2}{5}\right)$
is achieved.

\begin{figure*}[t]
\centering
\includegraphics[width=0.76\textwidth, keepaspectratio=true]{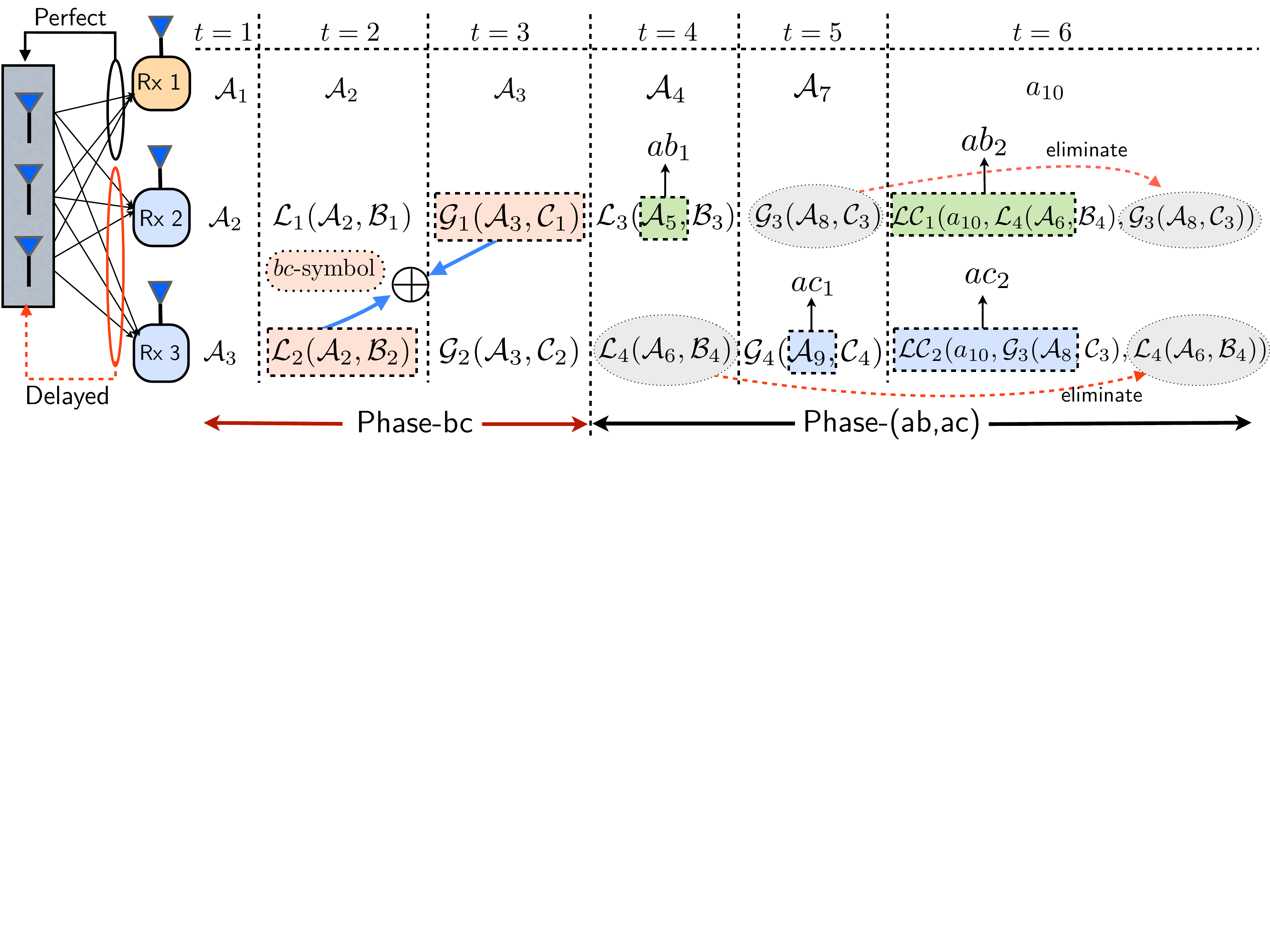}
\vspace{-155pt}
\caption{Generating order $2$ symbols -- $(3,3)$ MISO BC, $\CSIT$ configuration $\PDD$.}
\label{fig:3111_AllStages}
\vspace{-15pt}
\end{figure*}
\vspace{5pt}
\subsection{Theorem~\ref{Theorem3111PPD}:  $9/4$ $\DoF$ $-$ $(3,3)$ MISO BC $(\PPD)$}}
We present a scheme (shown in Fig.~\ref{fig:3111Stage1_Phase1_PPD}) that achieves the $\DoF$ triplet $(d_{1}, d_{2}, d_{3})= \left(1,1,\frac{1}{4}\right)$ i.e., $\DoF$ of $9/4$ in the $\PPD$ configuration for the $(3,3)$ MISO BC. In the $\PPD$ configuration, the transmitter
has instantaneous $\CSIT$ from the receivers $1,2$ and delayed $\CSIT$
from receiver $3$.  

At $t=1$, we send $2$ symbols each to the $1$st and $2$nd receivers $(a_{1}, a_{2})$, $(b_{1}, b_{2})$ in directions
orthogonal to the $\mathbf{h}_2(1)$ and $\mathbf{h}_1(1)$ respectively so as to not create cross interference. Additionally we transmit one symbol, $c$ for the $3$rd receiver by using a projection operator $\mathbf{h}_{(1, 2)}^{\perp}(1)$ such that 
$\mathbf{h}_1(1)\mathbf{h}_{1, 2}^{\perp}(1)=\mathbf{h}_2(1)\mathbf{h}_{1, 2}^{\perp}(1)=0$.
The transmitted signal is given by
\begin{align}
\mathbf{x}(1)=\mathbf{h}_2^{\perp}(1)\begin{bmatrix}
a_1\\
a_2\\
0
\end{bmatrix}+
\mathbf{h}_1^{\perp}(1)\begin{bmatrix}
b_1\\
b_2\\
0
\end{bmatrix}+
\mathbf{h}_{(1, 2)}^{\perp}(1)
\begin{bmatrix}
c\\
0\\
0
\end{bmatrix}.
\end{align}
This is repeated again at $t=2$ with new symbols $(a_{3}, a_{4})$ and $(b_{3}, b_{4})$ for $1$st and $2$nd receivers but with the same $c$ symbol for the $3$rd receiver as
\begin{align}
\mathbf{x}(2)=\mathbf{h}_2^{\perp}(2)\begin{bmatrix}
a_3\\
a_4\\
0
\end{bmatrix}+
\mathbf{h}_1^{\perp}(2)\begin{bmatrix}
b_{3}\\
b_{4}\\
0
\end{bmatrix}+
\mathbf{h}_{(1, 2)}^{\perp}(1)
\begin{bmatrix}
c\\
0\\
0
\end{bmatrix}.
\end{align}
At the end of these two time slots, notice from Fig.~\ref{fig:3111Stage1_Phase1_PPD} that receiver $1$ requires LCs $\mathcal{A}_2$ and $\mathcal{A}_4$. These symbols are also useful to the $3$rd receiver as it helps cancel interference and thereby decode symbol $c$. Similarly, $\mathcal{B}_2$ and $\mathcal{B}_4$ are required at the $2$nd and $3$rd receivers. 
Thus the goal of the next two time slots is to send these symbols in an efficient manner to the receivers. Using delayed $\CSIT$ from the $3$rd receiver,  transmitter can reconstruct $\mathcal{A}_2, \mathcal{A}_4,\mathcal{B}_2$ and $\mathcal{B}_4$.

At $t=3$, the transmitter sends the symbols $\mathcal{A}_2$ and $\mathcal{B}_4$ as
\begin{align}
\mathbf{x}(3)&= \mathbf{h}_2^{\perp}(3)[\mathcal{A}_2 \ \ 0  \ \ 0]^{T} +  \mathbf{h}_1^{\perp}(3)[ \mathcal{B}_4 \ \  0 \ \ 0]^{T}\nonumber
\end{align}
and the outputs at the receivers are shown in Fig.~\ref{fig:3111Stage1_Phase1_PPD}. Before we proceed further, let us summarize the transmission scheme until $t=3$. The $1$st receiver has LCs $\mathcal{A}_1,\mathcal{A}_2,\mathcal{A}_3$. The $2$nd receiver has the symbols
$\mathcal{B}_1,\mathcal{B}_3,\mathcal{B}_4$ and the $3$rd receiver has $\mathcal{L}_1(\mathcal{A}_2,\mathcal{B}_2,c)$, $\mathcal{L}_2(\mathcal{A}_4,\mathcal{B}_4,c)$ and $\mathcal{L}_3(\mathcal{A}_2,\mathcal{B}_4)$. The $1$st and $2$nd receivers need $\mathcal{A}_4$ and $\mathcal{B}_2$ respectively as it helps them decode their desired symbols. Hence, after $t=3$, receiver $3$ can:

\noindent $\bullet$ eliminate $\mathcal{A}_2$ from $(\mathcal{L}_1(\mathcal{A}_2,\mathcal{B}_2,c), \mathcal{L}_3(\mathcal{A}_2,\mathcal{B}_4))$ to form $\mathcal{L}_4(\mathcal{G}_1(\mathcal{B}_2,\mathcal{B}_4),c)$; and

\noindent $\bullet$ eliminate $\mathcal{B}_4$ from $(\mathcal{L}_2(\mathcal{A}_4,\mathcal{B}_4,c), \mathcal{L}_3(\mathcal{A}_2,\mathcal{B}_4))$ to form $\mathcal{L}_5(\mathcal{G}_2(\mathcal{A}_2,\mathcal{A}_4),c)$. 

At $t=4$, the transmitter sends 
\begin{align}
\mathbf{x}(4)\hspace{-2pt}=\hspace{-2pt} \mathbf{h}_2^{\perp}(4)[\mathcal{G}_2(\mathcal{A}_2,\mathcal{A}_4) \ \ 0  \ \ 0]^{T} \hspace{-2pt}+\hspace{-2pt}  \mathbf{h}_1^{\perp}(4)[ \mathcal{G}_1(\mathcal{B}_2,\mathcal{B}_4) \ \  0 \ \ 0]^{T}\nonumber,
\end{align}
%
From Fig.~\ref{fig:3111Stage1_Phase1_PPD}, it is clear that  receivers $1$ and $2$ can decode $(a_{1}, a_{2}, a_{3}, a_{4})$, and 
$(b_{1}, b_{2}, b_{3}, b_{4})$ respectively. Receiver $3$ can decode the symbol $c$ from $\mathcal{L}_4(\mathcal{G}_1(\mathcal{B}_2,\mathcal{B}_4),c)$, 
$\mathcal{L}_5(\mathcal{G}_2(\mathcal{A}_2,\mathcal{A}_4),c)$ and $\mathcal{L}_6(\mathcal{G}_1(\mathcal{B}_2,\mathcal{B}_4),\mathcal{G}_2(\mathcal{A}_2,\mathcal{A}_4))$. In summary, this scheme achieves the $\DoF$ triplet $(1,1,\frac{1}{4})$ or in other words the sum $\DoF$ of $\frac{9}{4}$. This concludes the achievability 
proof for Theorem~\ref{Theorem3111PPD}.

\section{Conclusions}\label{Conclusions}
In this paper, we investigated the impact of hybrid $\CSIT$ on the degrees-of-freedom of  $3$-receiver MISO BC. Novel achievable schemes were presented for various hybrid $\CSIT$ configurations which established the optimal $\DoF$ (for the $(2,3)$ MISO BC) and improved upon the best known achievable $\DoF$ (for the $(3,3)$ MISO BC). Our results show that that an important aspect when dealing with hybrid $\CSIT$ is the generation and transmission of higher order symbols which are desired by multiple receivers. As our schemes show, this problem is far from trivial even for the $3$-receiver broadcast channel. Showing the optimality of these schemes, and extensions of these ideas to more than $K=3$ receivers are interesting open problems for future work. 


\end{document}